\documentclass[twocolumn,aps,prb,showpacs]{revtex4}

\usepackage{graphicx}
\usepackage[usenames]{color}
\usepackage{amsmath}
\usepackage{amssymb}
\usepackage{subfigure}
\usepackage{pstricks}

\usepackage{dcolumn}
\usepackage{multirow}
\usepackage{float}

\def\ot{\frac{1}{3}}

\def\up{\uparrow}
\def\dn{\downarrow}
\def\ve{\varepsilon}

\def\unit#1{\ \mathrm{#1}}

\def\ci{e^2/(4\pi\ve\ell_0)}
\def\cunit{e^2/(4\pi\ve\ell_0)}

\begin{document}

\title{Transport gap in a $\nu=1/3$ quantum Hall system:
  a probe for skyrmions }

\author{Annelene F. Dethlefsen}
\author {Rolf J. Haug}
\affiliation{Institut f\"ur Festk\"orperphysik, Universit\"at Hannover,
Appelstra\ss{}e 2, D--30167 Hannover, Germany}
\author{Karel V\'yborn\'y}
\author{Ond\v rej \v Cert\'\i k}
\affiliation{Fyzik\'aln\'\i{} \'ustav, Akademie v\v ed \v Cesk\'e republiky,
  Cukrovarnick\'a 10,  CZ--16253 Praha 6, Czech Republic}
\affiliation{1. Institut f\"ur theoretische Physik, Universit\"at
  Hamburg, Jungiusstr. 9, D--22305 Hamburg, Germany}

\author{Arkadiusz W\'ojs}
\affiliation{Institute of Physics, Wroc{\l}aw University of
  Technology, Wybrze\.ze Wyspia\'nskiego 27, 50-370 Wroclaw, Poland}

\date{March 17th, 2006}


\begin{abstract}
The dependence of the activated gap on magnetic field is studied for
the fractional filling factor~1/3. By comparing the experimental
results with results from exact diagonalization calculations we are
able to identify the excitation of a small antiskyrmion in the
low-field regime and a cross-over to spinless excitations at higher
magnetic fields.  The effect of sample quality is studied. On the side
of the theory, comparison between different geometries (torus and
sphere) and different sizes is carried out. Under inclusion of Landau
level mixing and finite thickness, we obtain a good agreement between
calculated energies and experimental results.
\end{abstract}

\maketitle

\section{Introduction}

The existence of skyrmions is one of the remarkable many-body
phenomena accompanying the quantum Hall effects. After the
skyrmions  were
established\cite{barrett:06:1995,schmeller:12:1995,aifer:01:1996,maude:11:1996}
in the integer quantum Hall effect (IQHE) a question appeared
whether they can also be observed in the regime of the fractional
quantum Hall effect (FQHE), given the composite fermion (CF)
mapping between the IQHE and the FQHE\cite{heinonen:1998}. We
report here on an experiment indicating that the answer is
positive.

A skyrmion can be viewed as a finite-size quasiparticle of charge
$e$ located at $r_0$ in the parent ground state, which is e.g.
the fully polarized ($\up$) completely filled lowest Landau level
(LL). More precisely, it is the many-body ground state at magnetic
field $B$ corresponding to the filling factor $\nu=(n_eh/eB)=1$
minus one magnetic flux quantum ($n_e$ is the electronic
density, $h$ Planck's constant and $e$ the elementary charge of an
electron). In this state, called also a spin-texture, the
expectation value of spin is reversed ($\dn$) at $r=r_0$, it
remains unchanged ($\up$) for $r\to\infty$ and it interpolates
smoothly between $r_0$ and infinity\cite{fertig:10:1994}. A size,
$K$ (precise definition in Sec. \ref{pos-03}), can be attributed
to a skyrmion, related to how fast the spin changes with
displacement from the skyrmion center. For magnetic fields of
$\nu=1$ plus one magnetic flux quantum a symmetric quasiparticle
of charge $-e$ exists, an antiskyrmion.

The skyrmions at integer filling factors can either be studied using
Hartree-Fock\cite{fertig:10:1994} and field theoretic
methods\cite{sondhi:06:1993,lee:03:1990,moon:02:1995} on one side or
by exact diagonalization\cite{xie:01:1996,wojs:07:2002,rezayi:03:1991}
on the other side and all these approaches are
interrelated\cite{rezayi:09:1997}. The central conclusion is that
while Zeeman energy prefers small sizes $K$, meaning small average
number of reversed spins, the Coulomb (exchange) energy prefers
spatially smooth spin textures, i.e. large skyrmions, where two
neighbouring spins are almost parallel. The size of the skyrmion
lowest in energy is thus determined by the ratio of the Zeeman and
Coulomb energies, $\eta=E_Z/E_C=\mu_B g B/(\ci)\propto\sqrt{B}$, where
$\ell_0=\sqrt{\hbar/eB}$ is the magnetic length, $\mu_B$ the Bohr
magneton, $g$ the effective electronic Land\'e factor and
$\varepsilon$ the dielectric constant. This conclusion remains valid
also for integer filling factors\cite{wu:05:1995} $\nu>1$, albeit the
nonmonotonous Haldane pseudopotentials imply richer skyrmion
phase--diagrams\cite{wojs:07:2002}.

Works related to fractional filling
factors\cite{rezayi:10:1987,sondhi:06:1993,kamilla:??:1996,wojs:07:2002},
most importantly to $\nu=1/3$ which is the CF counterpart to $\nu=1$ of
electrons, lead to the same conclusion. However, apart of quantitative
differences in skyrmion energies, fractional and integer systems showed
some qualitative differences. The exact symmetry between skyrmions
($\nu>\ot$) and antiskyrmions ($\nu<\ot$) is absent\cite{wojs:07:2002} and
thus skyrmion and antiskyrmion sizes need not be the same in one system.
Also, temperature dependences of the magnetization are different in the
IQHE and FQHE regimes\cite{macdonald:10:1998,chakraborty:05:1996}.

Experimentally, the skyrmions were proven in magnetotransport and
in the Knight shift of the NMR or magnetoabsorption spectroscopy
sensitive to the spin polarization of the 2DEG.  The first method
probes skyrmion--antiskyrmion pairs as an excitation on the
background of the fully spin polarized (ferromagnetic) ground
state at exactly $\nu=1$. The other two methods probe the ground
state at a slightly changed filling factor. Provided the filling
factor is not too far from one, the ground state remains the
ferromagnetic state plus one skyrmion (antiskyrmion) per magnetic
flux removed (added) to the system. The depolarization in units of
electron spin per one magnetic flux is thus equal to the average
size of a skyrmion, i.e. to the number of involved spin flips.

From the transport activation gap at filling factor one, Schmeller
{\em et al.}\cite{schmeller:12:1995} concluded that a typical
excitation in a GaAs heterostructure contains seven spin flips. If the
excitation is a skyrmion--antiskyrmion pair, each of these should have
a size of $K=3$. The number of spin flips (size) was found to decrease
with increasing ratio of Zeeman and Coulomb energies $\eta$. The
optical experiments\cite{aifer:01:1996} and the NMR
experiments\cite{barrett:06:1995} gave approximately the same result.
Hydrostatic pressure reduces the effective Land\'e $g$-factor in GaAs
and even $g=0$ is experimentally possible.  It allows to access
smaller values of $\eta$ compared to performing measurements at
low magnetic fields where QHE will eventually disappear. Maude {\em
et al.}\cite{maude:11:1996} observed skyrmions as large as $K=16$ in
magnetotransport at nearly vanishing Zeeman energy.

The Coulomb energy stabilizing skyrmions is much smaller at
$\nu=\ot$ as compared to $\nu=1$. As a consequence, the skyrmions
in the FQHE regime are usually smaller. Leadley {\em et
al.}\cite{leadley:11:1997} found excitations with three spin flips
in magnetotransport at nearly $g=0$ implying skyrmion sizes
($K_S$) and antiskyrmion sizes ($K_A$) with $K_A+K_S+1=3$. In
contrast to this, NMR measurements by Khandelwal {\em et
al.}\cite{khandelwal:07:1998} suggested $K_A\approx K_S\approx
0.1$. The reason for this very different result is unclear.

Experimental arguments in favour of skyrmions at $\nu=\ot$ are by
far not so numerous as compared to integer filling factors. With
our new experiments agreeing well with exact diagonalization
calculations presented here, we believe, the existence of
skyrmions in the FQHE is confirmed as well as the
skyrmion-antiskyrmion asymmetry demonstrating the qualitative
differences between electrons and composite fermions.

To demonstrate the influence of the sample quality we present
measurements of the $\nu=\ot$ activation gap, $\Delta$, from two
gated heterostructures as a function of the electron density
$n_e$, i.e. the magnetic field $B$, Sec.~\ref{pos-02}. We observe
a linear $\Delta(B)$ behaviour in a large region of magnetic fields
implying that, roughly, the probed excitation costs much Zeeman
energy and little Coulomb energy. In particular, the lowest
excitation in the high mobility sample contains two spin flips
while it has a single spin flip for the low mobility sample. While
the latter case corresponds to a spin wave, a quasihole (QH) and a
quasielectron with reversed spin (QEr), the excitation with two
spin flips must contain either a skyrmion or an antiskyrmion.
Energies obtained by exact diagonalization (ED) support this
interpretation, Subsec.~\ref{pos-11}, and they indicate that we
observe a pair of the QEr and the smallest antiskyrmion, the one
containing one spin flip.

Moreover we observe a clear transition to another lowest excited
state for $B\gtrsim 9\unit{T}$ in the high mobility sample. The
energies obtained by the ED suggest that the relevant excitation
then contains no spin flip. This agrees with the usual statement
that such excitations in the high $B$ limit are the charge density
waves (CDW) with very large wavevector $k$. However, in the
present case we observe a remarkable coincidence between the
activation gap and the magnetoroton minimum, i.e. CDW at
$k\ell_0\approx 1.4$. We propose that this could be because the
activation is a two-step process where creation of a magnetoroton
is a bottleneck.

The exact diagonalization calculations, Sect.~\ref{pos-03}, start
with ideal systems (zero width, $\hbar\omega\gg \cunit$) and then
we take into account the finite thickness of the 2D electron gas
as well as the LL mixing. Assuming a $B$--field independent
reduction\cite{morf:09:2003,dethlefsen:08:2005} $E_d$ of the
activation gaps due to the disorder ubiquitous in experimental
samples, the calculated energies lead to a {\em quantitatively}
correct prediction of the activation gap with a single fitting
parameter~$E_d$.

\section{The experiment}
\label{pos-02}

The investigated two-dimensional electron systems are realized in
Al$_{0.33}$Ga$_{0.67}$As/GaAs heterostructures. The sample
growth-parameters are given in Tab.~\ref{tab-01}.
\begin{table}[h]
\begin{tabular}{|c|c|c|c|} \hline
\rule[-2mm]{0cm}{6mm}sample & spacer-&density (m$^{-2}$)& mobility (m$^2$/Vs)\\
                          &    width      &T~$\simeq~$40~mK, dark & T~$\simeq~$40~mK, dark \\
 \hline \hline
\rule[-2mm]{0cm}{6mm} \#1  & 70~nm       &$1.3 \cdot 10^{15}$& 700
\\ \hline
\rule[-2mm]{0cm}{6mm} \#2  & 40~nm       &$1.6 \cdot 10^{15}$& 79
\\ \hline
\end{tabular}
\caption{Parameters of the investigated samples.} \label{tab-01}
\end{table}
The basic difference between the samples is their mobility. The
high quality of sample~\#1 allows to study several different FQHE
states, whereas the mobility below $100\unit{m^2/Vs}$ is sufficient only
for studies of the most robust FQHE state, $\nu=1/3$, in
sample~\#2.

A metallic topgate enables us to vary the electron density in a
wide range. For sample~\#1 the electron density $n_{e}^{}$ is
varied between $0.2$ and $1.3 \cdot 10^{15}$ m$^{-2}$ with a zero
field mobility reaching $700$~m$^{2}$/Vs at 40~mK. With
sample~\#2, densities between $0.59$ and $1.6 \cdot
10^{15}$~m$^{-2}$ can be accessed. Here the zero field mobility reaches
$79$~m$^{2}$/Vs at 40~mK.

\begin{figure}[h]
\begin{center}
\includegraphics[width=8cm]{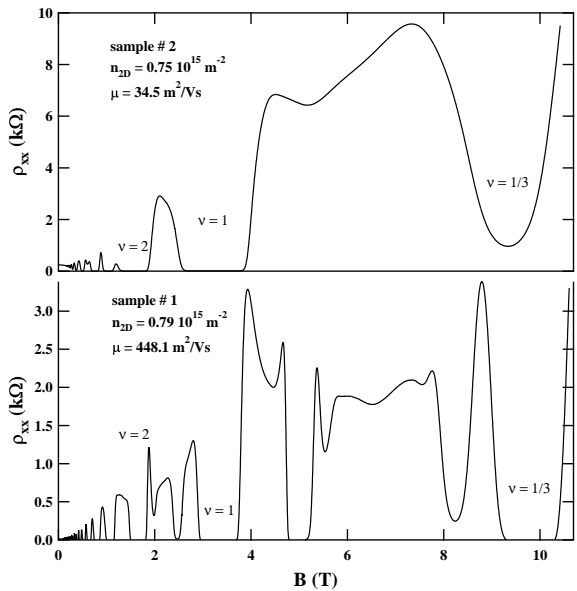}
\end{center}
\caption{Shubnikov--de Haas oscillations for both samples at
similar electronic densities.}\label{fig-10}
\end{figure}
The experiments are performed in a dilution refrigerator with
magnetic fields up to 20~T. The longitudinal resistivities
$\rho_{xx}$ (Shubnikov de-Haas oscillations) of the two samples at
nearly the same electron densities are shown in Fig.~\ref{fig-10}
demonstrating the different quality of the samples. While the
sample \#2 exhibits only one minimum between filling-factors
$\nu=1$ and $\nu=1/3$, for sample \#1 there is a series of
different fractional quantum Hall states in this region.

\begin{figure}[h]
\begin{center}\includegraphics[width=8cm]{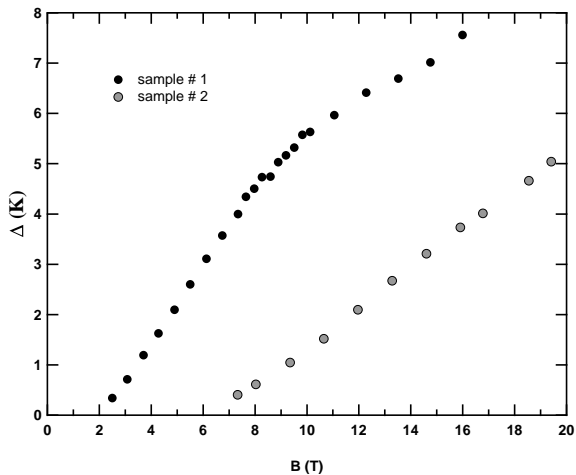}
\end{center}
\caption{Gap energies at $\nu=1/3$ from sample \#1 and \#2.} \label{Fig01}
\end{figure}

To obtain the activation energy for the different magnetic fields
we investigate the temperature dependence of the resistivity
minimum at $\nu=1/3$. The temperatures in our experiment are
varied between $T=40\unit{mK}$ and $1000\unit{mK}$. In this range
the error of measurement of the calibrated ruthenium oxide sensor
is $\pm 1$~mK. We extract the gap $\Delta$ out of
Arrhenius-plot data, using the activated resistance behaviour
$\rho_{xx} \propto \exp({-\Delta/2T})$. We assume that our total
uncertainty in $\Delta$ is less than 2\%. The measured activation
energies $\Delta$ are shown in Fig. \ref{Fig01} for the two
samples.

\section{Exact diagonalization}
\label{pos-03}

Electron-electron interaction is the fundamental effect giving
rise to most of the phenomena occurring within the lowest LL.
Because of the extremely high degeneracy of LLs,
standard techniques like Hartree-Fock approximation are
inapplicable for describing these phenomena.

In the exact diagonalization (ED) approach
\cite{chakraborty:1995,yoshioka:2002},
we start with the complete many-body Hamiltonian. It comprises of the
electron-electron interaction and the Zeeman energy.
\begin{eqnarray}\nonumber
  H &=& \sum_{J,\Sigma} {\cal A}_{J}
        c^\dag_{j_1\sigma_1}c^\dag_{j_2\sigma_2}
    c_{j_3\sigma_2}c_{j_4\sigma_1}
     + \sum_{j, \sigma} E_Z\frac{\sigma}{2} c^\dag_{j\sigma}c_{j\sigma}\\
    \label{eq-02}
     &&\qquad J=(j_1,j_2,j_3,j_4)\,,\quad \Sigma=(\sigma_1,\sigma_2)
\end{eqnarray}
Here $j_i$ is the orbital quantum number, $\sigma=\pm 1$ is the
spin and $c^\dag_{j\sigma}$ are the operators creating the
corresponding one-electron states. A convenient one-particle
orbital quantum number is either angular momentum or one component
of the linear momentum. These choices are typical for spherical
geometry and torus geometry, respectively.  The last two notions
describe the central approximation of the ED model. Instead of an
infinite plane, we study a compact manifold preferably without
edges, that is, we confine the electrons either to the surface of
a sphere \cite{haldane:08:1983} or to a rectangle with periodic
boundary conditions\cite{yoshioka:04:1983} (torus). The
Coulomb matrix elements ${\cal A}_J$ are given explicitly in
Reference 29 
for torus and they
straightforwardly follow from the Haldane pseudopotentials on a
sphere \cite{haldane:08:1983}.  The Zeeman energy is just
$E_Z=\mu_B g B$.

For the moment, we did not include any cyclotron energy
($\hbar\omega=\hbar eB/m^*$) term into
(\ref{eq-02}). All electrons are assumed to stay in the lowest LL
which is true for the ground state and low-lying excited states
if $\hbar\omega\gg {\cal A}_J, E_Z$ and $\nu\le 2$. This approximation,
exclusion of the LL mixing, is
justified in the high $B$ limit owing to $\hbar\omega, E_Z\propto B$,
${\cal A}_J\propto \sqrt{B}$ and $\hbar\omega/E_Z\approx 60\gg
1$.

The homogeneous magnetic field in the 2D system now corresponds to
$N_m=2Q$ quanta of magnetic flux passing through the surface of the sphere or
torus. In this situation, exactly $N_m$ one-electron states exist in
the system \cite{haldane:02:1985,chakraborty:1995}.
If we now put $N_e$ electrons into the system, the filling factor is
$\nu=N_e/N_m$ for the torus and $\nu=N_e/(N_m+\delta)$ for the
sphere. The quantity $\delta$ is of the order of unity ($\delta/N_m\to
0$ for $N_m\to\infty$) and it is related to the topology of the
considered eigenstate\cite{nayak:07:1995}.

The number of all possible $N_e$-electron states is then finite. The
matrix of Eq. (\ref{eq-02}) in this complete basis is evaluated and
diagonalized yielding the energies and many-body wavefunctions.

\subsection{Activation energy in transport}

It has been widely accepted that the activation gap $\Delta$ is the
lowest energy needed to create a neutral pair of charged particles and
to separate them very far from each
other\cite{chakraborty:??:1990,chakraborty:05:1990}.  Starting from
the Laughlin ground state at $\nu=\ot$, these particles are not an
electron and a hole. Rather they are particle-like many--body
excitations\cite{laughlin:05:1983} with charge $q$ and spin $s$.  They
are usually called quasielectrons (QE, $q=-|e|/3$, $s=1/2$) and
quasiholes (QH, $q=|e|/3$, $s=1/2$), eventually with reversed spin
(QEr, $q=-|e|/3$, $s=-1/2$) as regarding to the direction preferred by
the Zeeman energy. Creation energies of all these three quasiparticles
are different, even disregarding the Zeeman contribution, owing to
their actual many--body nature. This is a fundamental difference to
IQH systems.

Because of their charge $e/3$, all interactions between the mentioned
quasiparticles at $\nu=1/3$ (e.g. skyrmion energies discussed below)
are roughly weaker by an order of magnitude compared to $\nu=1$. The
interactions at long range are similar to the common Coulomb repulsion
or attraction because the size of the quasiparticles is of
the order of $\ell_0$. At short range, on the other hand, it's not
guaranteed that the interactions are similar to electrons because of
the internal structure of the quasiparticles\cite{wojs:01:2000,lee:08:2002}.  
Contrary to QE, the charge densities of QEr and QH are, however,
basically structureless and this fact lies at the heart of the close
analogy between low-energy excitations at $\nu=1/3$ and $\nu=1$.

\begin{figure}
\begin{center}
\includegraphics[scale=0.45]{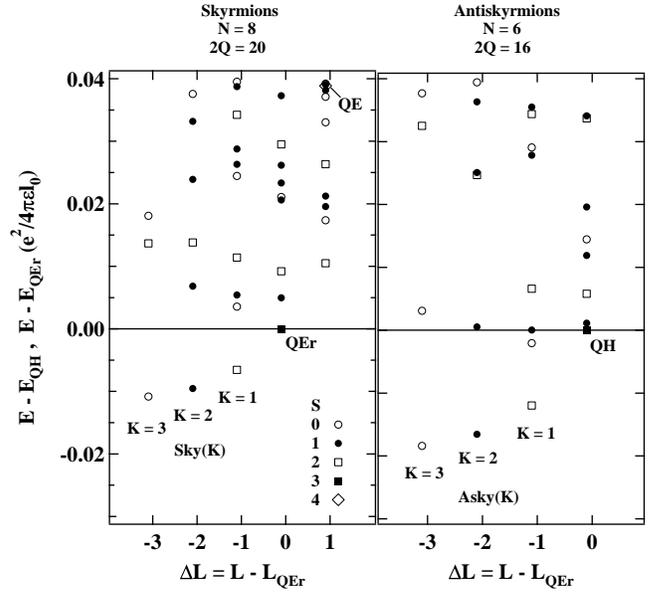}%
\end{center}
\caption{Full skyrmion and antiskyrmion spectra at $\nu=\ot$ minus and
  plus one magnetic flux quantum respectively and zero Zeeman energy.
  The (anti)skyrmion
  branches of $\Delta L=-K$ have negative energy and they
  are well separated from the continuum of
  excited states. For $\nu>\ot$,
  quasielectron with reversed spin (QEr) is lower in
  energy than QE. This holds for systems of all accessible sizes (here
  6 and 8 electrons).}
\label{fig-04}
\end{figure}

The Coulomb energy of skyrmions and antiskyrmions can be obtained from
the ED spectra {\em on a sphere}\cite{chakraborty:04:1997} in a
system with one flux quantum less and one flux quantum more,
respectively, than what would correspond to $\nu=\ot$,
Fig.~\ref{fig-04}. For $E_Z\gg {\cal A}_J$, the lowest excitation
to determine the activated transport will involve creation of a
QE+QH pair. As $E_Z$ decreases, the lowest excitation becomes
QEr+QH, because the energy of a QEr (Coulomb 'correlation' energy)
is lower\cite{wojs:07:2002} than the energy of a QE, 
Fig.~\ref{fig-04} left. This will however not remain true in the limit
$E_Z\to 0$. There are objects with even lower Coulomb energy than
QEr and QH. For each $K=1,2,\ldots$ there is one such object with
total spin $K+1/2$ and charge $-e/3$ and $e/3$. They are usually
called skyrmions (Sky) and antiskyrmions (ASky), respectively.
Contrary to IQH systems, energies of Sky($K$) and ASky($K$) are
different implying that the skyrmion size $K_S$ and the
antiskyrmion size $K_A$ need not be equal in the same system.
Sky($K$) [ASky($K$)] is the energy difference between the QEr [QH]
and the lowest state with angular momentum and spin
\begin{equation}
L=L_{\mathrm{QEr,QH}}-K\,,\qquad S=S_{\mathrm{QEr,QH}}-K.
\end{equation}
In a system of $N_e$ electrons, the angular momenta used are
$L_{\mathrm{QH}}=S_{\mathrm{QH}}=N_e/2$ and
$L_{\mathrm{QEr}}=S_{\mathrm{QEr}}=N_e/2-1$. With this definition,
ASky(0) is a QH and Sky(0) is a QEr, meaning that ASky($K$) contains $K$
flipped spins while Sky($K$) contains $K+1$ flipped spins.

\begin{figure}
\begin{tabular}{cc}
\includegraphics[scale=0.5]{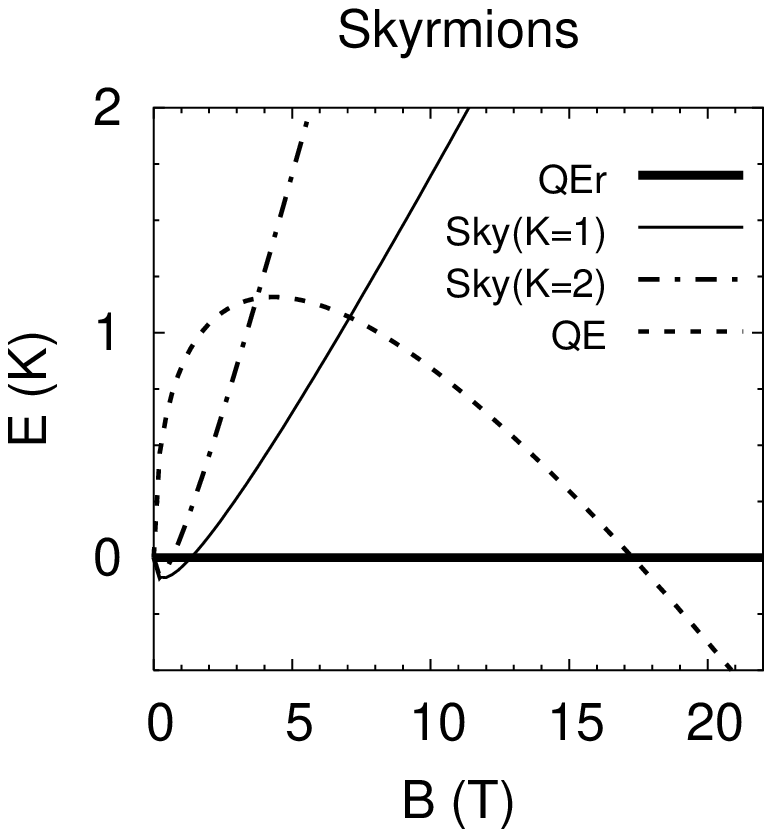} &
\hskip-5mm\includegraphics[scale=0.5]{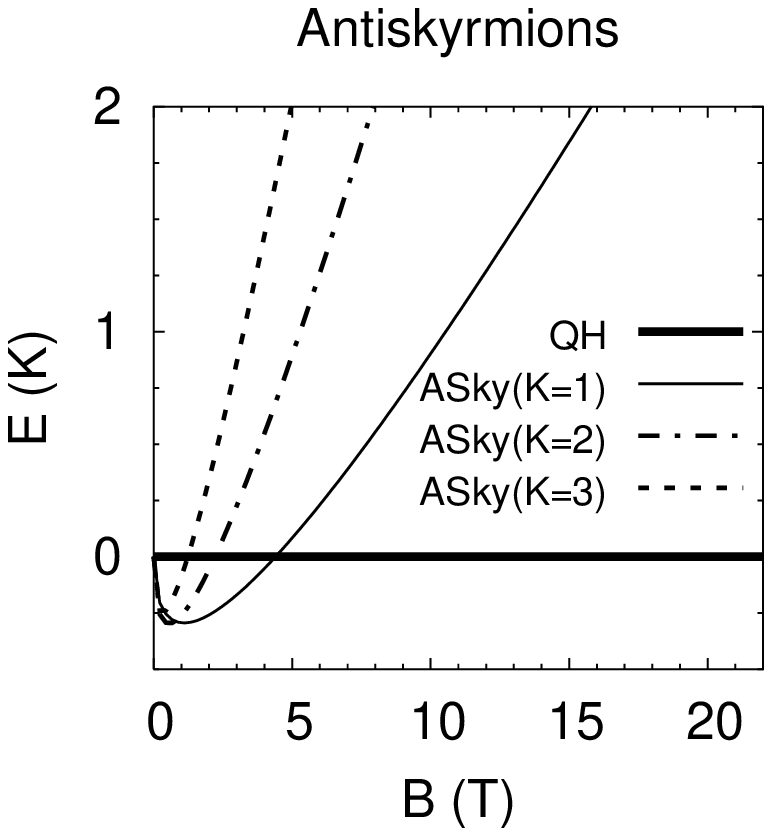}
\end{tabular}
\caption{Skyrmion and antiskyrmion energies at $\nu=\ot$ relative to
  QEr and QH in an ideal system with six electrons.
  The QE energy shown refers to the extrapolated $1/N\to 0$ values
  with finite thickness taken into account ($\beta=0.3$).}
\label{fig-03}
\end{figure}

Figure \ref{fig-03} shows the competition between skyrmions of
different sizes as a function of magnetic field, which means that the
ratio between the Coulomb and Zeeman energies, $\eta$, is varied. The
Zeeman energy prefers small skyrmions since these include less spin
flips. Thus, for fields above $4.5\unit{T}$ ($1.5\unit{T}$) no
antiskyrmions (skyrmions) occur in an ideal system as the Zeeman
energy $\propto B$ is then too large compared to the binding (Coulomb)
energy $\propto\sqrt{B}$. Owing to a rather high Coulomb energy cost,
the QE becomes favorable over QEr first at rather high magnetic
fields, Fig. \ref{fig-03} (left).

\begin{figure}
\begin{center}
\large
\includegraphics[scale=0.63]{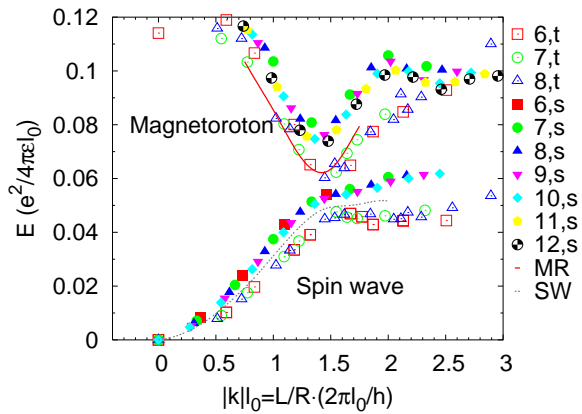}
\end{center}
\caption{The spin wave (SW) and the magnetoroton branch (MR) seen in
  the ED spectra of ideal $\nu=\ot$ systems of different sizes and
  geometries. In the legend, 't' stands for torus, 's' for sphere and
  the number indicates the number of electrons. The lines (solid and
  dotted) were obtained from the $1/N\to 0$ extrapolation of the data (MR
  and SW) on the sphere.}
\label{fig-08}
\end{figure}

Once a neutral pair of quasiparticles, Sky($K_S$) and ASky($K_A$), has
been created, they behave similarly to a magnetoexciton. In a magnetic
field, the magnetoexciton has a constant linear momentum $k$ which is
proportional to the mutual distance $\Delta x$ between the
quasiparticles. In the classical approach we would expect its energy
to be $E(\Delta x)\propto 1/\Delta x$ with proportionality constant
determined by the charges of the two constituent quasiparticles. In
the ED spectra such modes can be identified, Fig.~\ref{fig-08}.  They
are usually called magnetoroton (MR) branch, $E_{CDW}(k)$, for QE+QH
and spin wave (SW), $E_{SW}(k)$, for QEr+QH. The limiting values for
$k\to\infty$ are the energies necessary to create a QE+QH (QEr+QH)
pair and to separate them far from each other. These are the quantities
commonly used for comparison to the transport activation gaps, because
the SW (MR) is the lowest excitation (at $k\gtrsim 1.0\ell^{-1}$)
among all states with total spin $S=N/2-1$ ($S=N/2$), i.e. with one
(with no) spin flip.

It is remarkable how much $E_{CDW}(k)$ calculated on a sphere and
on a torus differ, on a quantitative level, Fig.~\ref{fig-08}.
Even though the positions of the magnetoroton minimum match well
in both geometries ($k\ell_0\approx 1.4$), sphere gives seemingly
a higher energy of the minimum by as much as 20\%. A careful
extrapolation to infinite systems (solid line in Fig.
\ref{fig-08}) however matches excellently the results obtained on
a torus. This is not surprising, given the magnetoexcitonic
character of the MR. The MR of $\Delta x$ comparable to the radius
of the sphere will have the QE and the QH located near the
opposite poles. Such situation is not compatible with a picture of
a plane wave of $k=\Delta x/\ell_0^2$ propagating along the
equator. On the other hand, with increasing radius of the sphere
$R$ this becomes a finite size effect if $R\gg \Delta x$. Based on
Fig.~\ref{fig-08}, we believe finite--system data from the torus
are more suitable to give quantitative estimates for magnetoroton
and spin wave energies.

For a Sky($K_S$)-ASky($K_A$) pair, we take $E_{SW}(k)$ with
$k\to\infty$ and add the creation energies of Sky($K_S$) and of
ASky($K_A$). Instead of one system, as it was the case for
studying the QEr+QH pair, we thus have to exactly diagonalize
three different systems: one for the quasiparticle-separation
procedure, one for the Sky and one for the ASky. This more
complicated procedure suffers possibly less from the finite size
effects, since skyrmions are rather extended objects, in
particular more extended than a bare QH or QEr. Recall that sizes
of Sky and ASky need not be the same.

\subsection{Finite thickness and LL mixing}

Aiming at the description of experiments under realistic conditions,
three ever--valid facts should not be left unnoticed: the sample is
actually three--dimensional (finite extent of the wavefunction
perpendicular to the 2DEG), the magnetic field is finite (mixing
between Landau levels) and the system is never perfectly homogeneous
(disorder).

Non-zero thickness $w$ of the 2DEG can be effectively incorporated into
the Haldane pseudopotentials\cite{haldane:08:1983} which completely
determine the Hamiltonian of the lowest LL. Qualitatively,
the larger the effective thickness $w/\ell_0$, the more softened
becomes the effective electron-electron interaction on shortest
distances.

Quantitative effects of the presence of the third dimension have been
studied since the early times of the FQHE, both with the Laughlin
state\cite{macdonald:05:1984} and the activation
gap\cite{zhang:02:1986}. In a heterostructure, electrons are confined
to a nearly triangular potential well. A standard choice for the
wavefunction in the growth direction is then the Fang--Howard trial
wavefunction\cite{ando:04:1982}, $\psi_{FH}(z)=(b^3/2)^{1/2}\
ze^{-bz/2}$.  We will mostly stay with this choice, even though we are
aware of other options for $\psi(z)$ which may lead to slightly lower
subband energies (Sec. V in Morf {\em et al.}\cite{morf:08:2002}).
Differences originating from these different choices of $\psi(z)$
should be smaller than the uncertainity in the variational parameter
$b$ (or the thickness of the 2DEG) relevant for our experiments. This
has been checked with $\psi_{QW}(z)=\cos \pi az$, $|z|<\pi/2$, relevant
for symmetric quantum wells. Taking $\psi_{FH}(z)$ instead of
$\delta(z)$ is equivalent\cite{zhang:02:1986}  to using
a nontrivial form--factor $F(q)$ in the 2D Fourier transforms $V(q)$ of the
Coulomb interaction
\begin{equation}\label{eq-03}
  V(q)=\frac{F(q)}{q}\,,\ F(q)=\frac{8+9(q/b)+3(q/b)^2}{(2+2q/b)^3}\,.
\end{equation}
The quantity $V(q)$ then enters the Coulomb matrix elements in
(\ref{eq-02}) as given in standard
references\cite{yoshioka:03:1986,chakraborty:1995}. These can be in
turn reexpressed in terms of the Haldane
pseudopotentials\cite{vyborny:2005} $V_m$. For reasonable values of
$b$, only $V_0$ changes appreciably, it decreases by 25\% for
$b^{-1}=0.3\ell_0$.

Spatial extent of the wavefunction along $z$ defined as FWHM is
$w\approx 4.9/b$ for $\psi_{FH}$ and $w=\frac{2}{3}/a$ for $\psi_{QW}$.
The wavefunction parameter $b$ depends on the form
(steepness) of the triangular well potential and therefore it is
not constant but it changes with the applied gate voltage. This leads
to \cite{zhang:02:1986,ando:04:1982}
\begin{equation}\label{eq-04}
  b=[33\pi m^* e^2 n_{e}/2\ve \hbar^2]^{1/3}\,,
\end{equation}
which depends only on the electron density\cite{chakraborty:1995}
$n_{e}$ and the dielectric constant $\ve$. If we assume the filling factor
fixed to $1/3$, the density becomes a function of the magnetic field,
so that
\begin{equation}\label{eq-05}
  \beta=(b\ell_0)^{-1}\approx 0.23\times (B[\mathrm{T}])^{1/6}\,.
\end{equation}
This is a formula relevant for both our samples.

The {\em LL mixing} is more difficult to include. If we admit that
higher Landau levels may also be populated even at $\nu<1$, we must
(i) add the cyclotron energy term $\sum_{nj\sigma} (n+1/2)\hbar\omega
c^\dag_n c_n$ to the Hamiltonian (\ref{eq-02}). We also have to
considerably extend the many-body basis (ii) because we have
introduced a new single-particle orbital quantum number, the Landau
level index $n$.  The former fact also implies that we have a new
energy scale $\propto B$ in the problem. Recall here the criterion for
the neglect of LL mixing: $1\gg {\cal A}_J/\hbar\omega\propto
1/\sqrt{B}$. Fortunatelly, the magnetic fields relevant for the FQHE
are still high enough for LL mixing to be treated perturbatively. In
practice this means, that in the first (second) order we allow for
maximum one (two) particles to be in the first LL ($n=1$) when constructing
the many-body basis.  For the current purpose we allowed for up to two
particles in the first Landau level and verified in small systems that
increasing this number does not change the energies perceptibly.

Without higher LLs, the energies $E_C$ of Hamiltonian's
(\ref{eq-02}) Coulomb part were conveniently evaluated in the
Coulomb units $e^2/(4\pi\ve \ell_0)$. Then the energies were
magnetic-field-independent for $E_Z\equiv 0$ and depended via
$S_z$ trivially on $B$ for $E_Z\not= 0$, in particular
$E_Z/(\cunit) \propto S_z\sqrt{B}$. With other LLs included in
addition to the lowest one, $E_C/(\cunit)$ becomes a function of
$B$ or better of\cite{yoshioka:03:1986}
$\lambda=\hbar\omega/(\cunit)$. However, since variations of
$E_C/(\cunit)$ as a function of $B$ ($\lambda$) are typically
small, see Fig. \ref{fig-09}, we will adhere to the Coulomb units.

{\em Disorder} is to the best of our knowledge the only relevant
effect {\em not} described microscopically within this work. A
common notion is that the disorder reduces the incompressibility
gap\cite{willett:05:1988}. In fact, randomly distributed potential
impurities included into the system (\ref{eq-02}) change energies
of both the ground state and the excited state. Because the
excited states in question consist of two microscopic
quasiparticles on the background of the Laughlin ground state, we
will assume for our purposes that the disorder indeed reduces the
excitation energy by a
constant\cite{morf:09:2003,dethlefsen:08:2005}.  The reduction
$E_d$ is typically of the order of one Kelvin. It need not be the
same for different excitations (skyrmions, magnetorotons) and this
causes probably the largest uncertainity in the respective
energies of spin textures of different sizes. Values of magnetic
field where transitions occur remain the same as in the disorder-free
systems, Fig.~\ref{fig-03}, only as long as the gap reduction is
equal for the two involved excitations.

\subsection{Numerical data}
\label{pos-11}

\begin{figure}
\begin{center}
\includegraphics[scale=0.6]{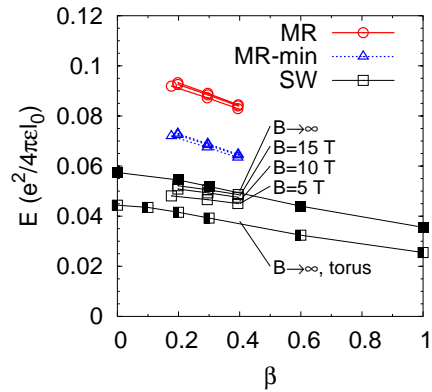}
\end{center}
\caption{The effect of the finite thickness and the LL mixing on
the magnetoroton (MR) minimum and on the best approximation to the
$k\to\infty$ energies of the spin wave (SW) and the MR branch.
Finite  system (seven electrons) on a sphere is considered, except
the single  curve marked 'torus'. Energies labeled with $B$
include the Landau  level mixing which vanishes for $B\to\infty$.}
\label{fig-09}
\end{figure}

Finite thickness of the 2DEG and LL mixing do not change the SW
and MR dispersions qualitatively. Their effect is that the ideal
dispersions $E_{CDW}(k)$ and $E_{SW}(k)$ become multiplied by a nearly
$k$--independent coefficient. The limiting values for $k\to\infty$
are reduced, both for nonzero thickness and LL mixing, Fig.
\ref{fig-09}. For experimentally relevant values of the
parameters, finite thickness decreases $E_{SW}(k\to\infty)$ by
10\% (for $\beta=0\to 0.3$) and the LL mixing up to the second
order decreases $E_{SW}(k\to\infty)$ by 10\% (at $B=5\unit{T}$).

Our best estimate for $E_{SW}(k\to\infty)$ starts with the torus.
It includes finite thickness ($\beta=0.3$) and LL mixing (as of
$B=5\unit{T}$) and it reads $0.035\ci$. Fig. \ref{fig-08} suggests
that this value almost would not change if we studied larger
systems. Note that with increasing magnetic field $\beta$ of the
heterojunction increases, Eq. (\ref{eq-05}), while the LL mixing becomes
less important. Quantitatively, the latter effect is stronger so
that the indicated value of $E_{SW}(k\to\infty)$ in Coulomb units
will slightly increase with increasing magnetic field, cf. Fig.
\ref{fig-09}.

\begin{figure}[h]
\begin{center}
\includegraphics[scale=0.6]{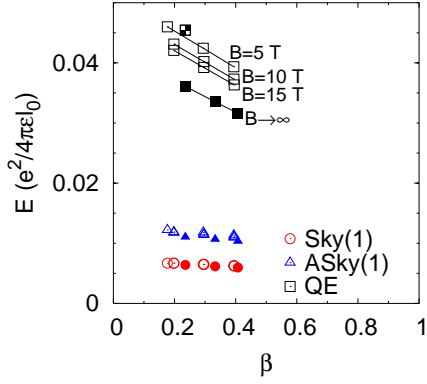}
\end{center}
\caption{Analogous to Fig. \ref{fig-09} but for the QE energy
relative to the energy of QEr as well as for energies of the
smallest skyrmion and antiskyrmion. All data were obtained on a
sphere, solid symbols refer to energies under no LL mixing.} \label{fig-06}
\end{figure}

\begin{table}[h]
\begin{tabular}{|c|c|c|c|c|c|c|}
\hline
       &\multicolumn{2}{|c|}{\large \bf torus}&\multicolumn{4}{|c|}{\large \bf sphere}\\
\hline
        &       & finite   &           &finite     &  + LL     &+ tdyn              \\
       &       &  width    &           &width      &  mixing    &limit              \\
       &ideal  &$\beta=0.3$& ideal     &$\beta=0.3$&  B=5T    &$1/N\to 0$         \\
\hline
    SW &0.045  &0.039      &   0.057   &   0.052   &   0.047  &                    \\
    MR &0.093  &0.080      &   0.102   &   0.089   &   0.087  &                    \\
    MR-min&0.063&0.054     &   0.076   &   0.069   &   0.067  &                    \\
\hline
    Sky(1) &    &          &   -0.0062 &   -0.0056 &   -0.0065&    -0.0050         \\
    ASky(1)&    &          &   -0.0112 &   -0.0102 &   -0.0118&    -0.0088         \\
    QE     &    &          &   0.0385  &   0.0335  &    0.0424&     0.0222         \\
\hline

\end{tabular}
\caption{Energies in $\cunit$ concerning Fig. \ref{fig-09} (SW, MR, MR-min) and
Fig. \ref{fig-06} (Sky(1), ASky(1), QE).} \label{tab-02}
\end{table}

\begin{figure} [h]
\begin{center}
\includegraphics[scale=0.6]{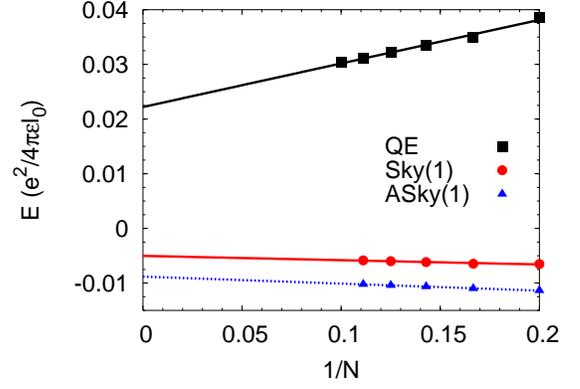}
\end{center}
\caption{Extrapolation of the (anti)skyrmion and QE energies to
the thermodynamical limit $1/N_e\to
  0$. Sphere, no LL mixing, finite width $\beta=0.33$.}
\label{fig-07}
\end{figure}

The skyrmion and antiskyrmion spectra at $\nu=\ot$ have been
introduced in the previous paragraph, Fig. \ref{fig-04}. At the
magnetic fields of our experiment only the smallest
(anti)skyrmions are possible, Fig. \ref{fig-03}. Contrary to the
spin wave: Their condensation energies relative to a bare QEr and
QH, show a slight dependence on
the system size, Fig. \ref{fig-07}. A linear fit in $1/N$ leads to
(anti)skyrmion energies by about 10\% lower at $1/N\to 0$ than
they are for $N_e=7$, Fig. \ref{fig-06} and Table~\ref{tab-02}. In
contrast to this, the QE energy becomes reduced by as much as
35\%.

The finite thickness ($\beta=0.3$) also reduces the Sky(1),
ASky(1) and QE energies by about 10\%, 10\% and 20\%,
respectively. On the
other hand, the Landau level mixing (1st order, $B=5\unit{T}$)
increases these energies by 20\%, 15\% and 25\%, respectively,
Fig. \ref{fig-06} and Table~\ref{tab-02}. 
The best guesses for the energies in $\ci$
relevant to our experiment is the following: $-0.0058$ for Sky(1)
and $-0.0102$ for ASky(1) for $B=5\unit{T}$ and a thickness
corresponding to $\beta=0.3$ and $0.0240$ for QE at $B=15\unit{T}$
and $\beta=0.4$. All these energies will definitely be reduced
when magnetic field is swept up because both LL mixing and finite
thickness have this tendency.

The data extracted from Fig. \ref{fig-09} and \ref{fig-06} are
summarized in Table \ref{tab-02}.

\section{Interpretation of the experimental data}
\label{pos-04}

\begin{figure}[h]
\includegraphics[scale=0.7]{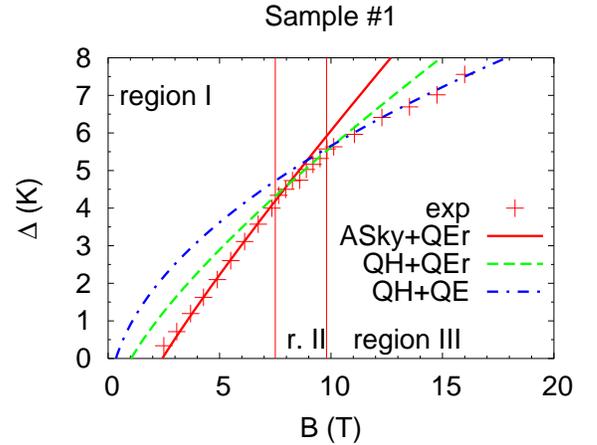}
\caption{Gap energies of the sample \#1 (Fig. \ref{Fig01})
interpreted as an antiskyrmion plus QEr for low $B$ (region I) and
a QE--QH pair for high $B$ (region III). The disorder-induced gap
reduction ($E_d$) obtained by fitting is $3.2\unit{K}$,
$1.8\unit{K}$ and $1.4\unit{K}$ for region I, II and III,
respectively.} \label{fig-01}
\end{figure}

The $B$--dependence of the activation gap in the {\em sample} \# 1 shows
an apparent transition slightly below $10\unit{T}$. Let us divide the
investigated range of magnetic field into three regions as shown in Fig.
\ref{fig-01}: I (low field), II (transition) and III (high field).

In the following we wish to argue that the lowest excitation which determines
the activation gap in region I is a QEr--antiskyrmion ($K_A=1$) pair. In region
III, QE--QH pairs without spin flip are observed while the QEr--QH pairs
likely show up in region II.

Our discussion begins with a general observation that the gap can change either
$\propto B$ (Zeeman energy)  or $\propto \sqrt{B}$ (Coulomb energy)
\begin{equation}\label{eq-01}
  \Delta [\!\unit{K}] 
  =E_C\cdot 50.2 \sqrt{B[\!\!\unit{T}]} +
  \Delta (S_z/\hbar)\cdot 0.295 B[\!\!\unit{T}]-E_d\,.
\end{equation}
Here the Coulomb energy $E_C$ should be put in units $\ci$ and
$\Delta(S_z/\hbar)$ means the number of spins flipped in the
excitation.

Let us first focus on the low field region I, Fig.~\ref{fig-01}. 
The experimental data show a remarkably precise
linear behaviour. Within the uncertainity of the measurement, this
linearity does {\em not} necessarily mean that the Coulomb
contribution to the gap entirely vanishes, but it sets a rather
stringent upper limit of approximatelly $E_C<0.025$. We are then
left merely with two possibilities: either $\Delta(S_z/\hbar)=3$
and $E_C$ is zero or $\Delta(S_z/\hbar)=2$ and $E_C\approx 0.02$.
The exact diagonalization spectra indicate that for
$B<4.5\unit{T}$ a $K_A=1$ antiskyrmion is energetically more
favourable than a bare quasihole or a ASky(2), Fig. \ref{fig-03}.
On the other hand, even the smallest skyrmion $K_S=1$ costs more
energy in this range of magnetic fields than a quasielectron with
reversed spin. Therefore, in this case, the most likely pair of
charged particles created by a thermal excitation will be a
QEr--ASky ($K_A=1$). Energetical cost of this excitation, two spin
flips plus Coulomb energy $E_C\approx 0.035-0.011=0.024$, Subsect.
\ref{pos-11}, is in a nice agreement with the experimental data,
Fig. \ref{fig-01}.

As the magnetic field increases, the $K_A=1$ antiskyrmion becomes more
energetically costly than a plain quasihole. The gap should then
amount to creation of a QH--QEr pair, i.e. to one spin flip plus
$E_C\approx 0.035$. It is not guaranteed that we indeed
observe this excitation in the measurement: region II is not very
large, Fig.~\ref{fig-01}, and it could well be, that what we observe here
is just a smooth transition between regions I and III. Despite this,
we should like to point out, that the exact diagonalization energies
are compatible with the QH--QEr interpretation of region~II.

Finally, for yet higher magnetic fields, it is more favourable to
create a quasielectron in a higher CF LL than to flip its spin. In
line with the situation of the $B\to\infty$ limit, we expect a
QH--QE pair without reversed spin to be the lowest excitation,
Fig. \ref{fig-03}. Indeed, the best fit in region III has
$E_C\approx 0.045\pm 0.005$ and zero Zeeman energy. This Coulomb
energy is however almost by a factor of 2 smaller than
$E_{CDW}(k\to\infty)$ from the exact diagonalization, Fig.
\ref{fig-09} and table \ref{tab-02}. Therefore, let us focus on
region III now.

The only alternative interpretation of region III is that we
observe a QH--QEr pair here (while region II is just the smooth
transition between the two other adjacent regions). This scenario is
not very likely though. With one spin flip included, Eq. (\ref{eq-01})
implies that the Coulomb energy in region III could not exceed
$0.010$. There is no justification for such an excitation within the
ED spectra in the $S=N/2-1$ sector, Fig. \ref{fig-08}. Also,
with this scenario, the gap reduction constant would be negative
(around $-1\unit{K}$).

With the interpretation of $E_C\approx 0.045$, $\Delta S_z=0$ it is
remarkable how near this value lies to the energy of the magnetoroton
minimum, Fig. \ref{fig-09} (Table \ref{tab-02}). It is tempting to
conclude that the activation process goes in two steps, creation of a
magnetoroton and unbinding of the constituent QE and QH. The former
step costs more energy, roughly $E_{CDW}(1.4\ell_0^{-1})=0.05$ in
Coulomb units as compared to
$E_{CDW}(k\to\infty)-E_{CDW}(1.4\ell_0^{-1})=0.08-0.05=0.03$ for the
unbinding of a magnetoroton. The creation is therefore a bottleneck
for the whole activation process and it determines the activation
energy measured in transport in the limit of high Zeeman
energies. Such a two-step process is not possible for the spin-flip
excitations because the spin wave dispersion has no minimum which
could lead to a stable intermediate state. We wish to stress that the
activation gap smaller than the theoretical predictions of
$E_{CDW}(k\to\infty)$ has been observed many
times\cite{willett:05:1988,boebinger:10:1985,boebinger:11:1987} but
the problem was never conclusively resolved. Usually, this discrepancy
was as whole attributed to the disorder. Here we propose that the
smaller observed gap for no-spinflip excitations is only in part due
to the disorder. For the sample~\#1, this $B$-independent reduction is
$E_d\approx 1.4\unit{K}$. The other, 'traditional', interpretation
that the disorder reduces the Coulomb energy, meaning $E_d=0$ and a
modified value of $E_C$ in (\ref{eq-01}), is in conflict with the gaps
observed in region I, Fig.~\ref{fig-01}. It should be noted that we
indeed found different gap reductions for different excitations within the
same sample, Fig.~\ref{fig-01}. An attempt to use the value of $E_d$
related to ASky(1)+QEr (region I) also for region III leads to only a
slightly changed $E_C$ while the quality of the fit is apparently worse.

\begin{figure}
\includegraphics[scale=0.7]{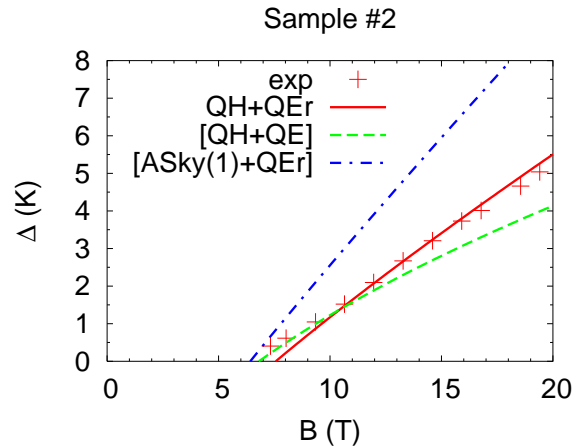}
\caption{Gap energies of the sample \#2 (Fig. \ref{Fig01})
interpreted as a spinwave (a QH--QEr pair). The gap reduction
$5.8\unit{K}$ is larger than for the sample \#1. For other options
(2 spin flips = ASky(1)+QEr and 0 spin flips = QE+QH), the
parameters were taken as in Fig. \ref{fig-01}, only the constant
offset $E_d$ was adjusted.} \label{fig-02}
\end{figure}

The data of {\em sample} \#2 suggest that we measure a QH--QEr pair in
the whole range of accessible magnetic fields, Fig.~\ref{fig-02}.  The
ED energy of a spin--wave is somewhat larger than what the
experimental data suggest ($0.025$). Other explanations, however, are
unlikely. Excitations with two spin flips (e.g. skyrmions) lie well
above the measured gaps, even if we assume zero Coulomb energy of suchan excitation. Spinless excitations, on the other hand, lead to
$\Delta(B)$ which is too far away from a linear dependence seen in
Fig.~\ref{fig-02}. As an example we took the two--spinflip (ASky+QEr) and
zero--spinflip (QH+QE) excitations as discussed for sample~\#1,
adjusted $E_d$ and plotted them into Fig.~\ref{fig-02} as a
dash--dotted and dashed line, respectively.

The absence of skyrmionic excitations for sample~\#2 is not
surprising given its lower mobility. Lower quality means larger
disorder--induced gap reduction ($E_d \approx 6\unit{K}$) implying
a higher FQHE threshold in $B$ ($B\gtrsim 7\unit{T}$), cf.
Fig.~\ref{fig-02} and~\ref{fig-01}. These are too high fields for
(anti)skyrmions to be observed, Fig.~\ref{fig-03}. Less obvious is
the absence of a transition to a spinless excitation (QE+QH) as
the one observed for sample~\#1. We find, however, that such an
excitation would be observable below $20\unit{T}$ only if
$E_d$ for QE+QH were $>5\unit{K}$. By comparison with typical gap reductions in
sample~\#1 this seems unlikely.

The present measurements suggest that, paradoxically, single spin-flip
excitations may be observed up to rather high magnetic fields
($20\unit{T}$) even in samples with mobility below $100\unit{m^2/Vs}$.
However, in order to observe larger (anti)skyrmions in FQH systems the
Zeeman energy should be suppressed\cite{kukushkin:07:1999}.  By applying the
hydrostatic pressure and reducing the Land\'e $g$--factor, the maximum
of three spin flips per excitation was reached\cite{leadley:11:1997}
compared to two spin flips of our experiment. 
In an ideal case, one should be
able to observe more transitions in $\Delta(B)$, not just one as in
Fig.~\ref{fig-01}, corresponding to successive reduction of skyrmion
and antiskyrmion sizes with increasing magnetic field (or Zeeman
energy) at fixed filling factor.

\section{Conclusion}

Spin excitations in the $\nu=\ot$ FQH system were studied using
measurements of the activation gap, $\Delta$, as a function of
magnetic field. Supported by energies obtained by exact
diagonalization we identified the activation-relevant excitation
to be a spin wave in the sample~\#2 and an antiskyrmion with one
spin flip plus a quasielectron with reversed spin for the
sample~\#1. The abrupt change in $\Delta(B)$ observed at $B\approx
9\unit{T}$ in the sample~\#1 was attributed to the transition to a
charge density wave in the lowest excitation. Since the gap was in
this case smaller than what we would expect for a charge density
wave with infinite wavevector, we proposed that the activation is
a two-step process with magnetoroton minimum governing the
activation energy as a bottleneck. With this interpretation, we
found the effect of disorder to be a constant reduction $E_d$ of
the gap, independent on magnetic field, in agreement with previous
works\cite{morf:09:2003,dethlefsen:08:2005}. Consistent with its
lower mobility, the gap reduction is larger for sample~\#2 and it
is different for different types of excitations.

In order to obtain a quantitative agreement between the
energies from the exact diagonalization and the experiment,
finite thickness as well as the Landau level mixing up to
the first order have to be included. We wish to stress that the number of spin
flips involved in the particular excitations can be determined with
very high certainty even with little knowledge of the Coulomb energy.
This is on one hand owing to the precision of the
experimental data showing linear $\Delta(B)$ and on the other hand
because the number of spin flips should be an integer.
Our only fitting parameter was the constant disorder--induced
reduction of the activation gap.

\medskip

\section*{Acknowledgements}

It is our pleasure to thank Eros Mariani and Daniela Pfannkuche for
worthwile discussions. AD and RH acknowledge support via DFG priority
program "Quanten Hall Systeme" and BMBF, AW acknowledges support from
the grant 2P03B02424 of the Polish MENiS and KV and O\v C acknowledge
the support by the Academy of Sciences of the Czech Republic and its
Grant Agency under Institutional Support No. AV0Z10100521 and by the
Ministry of Education of the Czech Republic Center for Fundamental
Research LC510.

\bibliographystyle{apsrev}

\end{document}